\documentclass[runningheads]{llncs}
\usepackage[utf8]{inputenc}

\usepackage{amssymb,amsmath}
\usepackage{graphicx}

\usepackage{pifont,nicefrac,euflag}
\usepackage[clock]{ifsym}
\usepackage{todonotes}
\usepackage[normalem]{ulem}
\usepackage{booktabs,array}
\usepackage[T1]{fontenc}
\usepackage{listings}
\usepackage{color}
\definecolor{brown}{rgb}{0.7,0.3,0.2}
\definecolor{darkgreen}{rgb}{0.0,0.2,0.0}
\definecolor{darkred}{rgb}{0.5,0.1,0.1}
\definecolor{gray40}{rgb}{0.4,0.4,0.4}
\definecolor{gray50}{rgb}{0.5,0.5,0.5}

\usepackage[colorlinks, urlcolor=blue, linkcolor=blue, citecolor=blue]{hyperref}
\usepackage{tablefootnote}

\newsavebox\euflagbox

\begin{document}

\setbox\euflagbox\hbox{\includegraphics[scale=0.04]{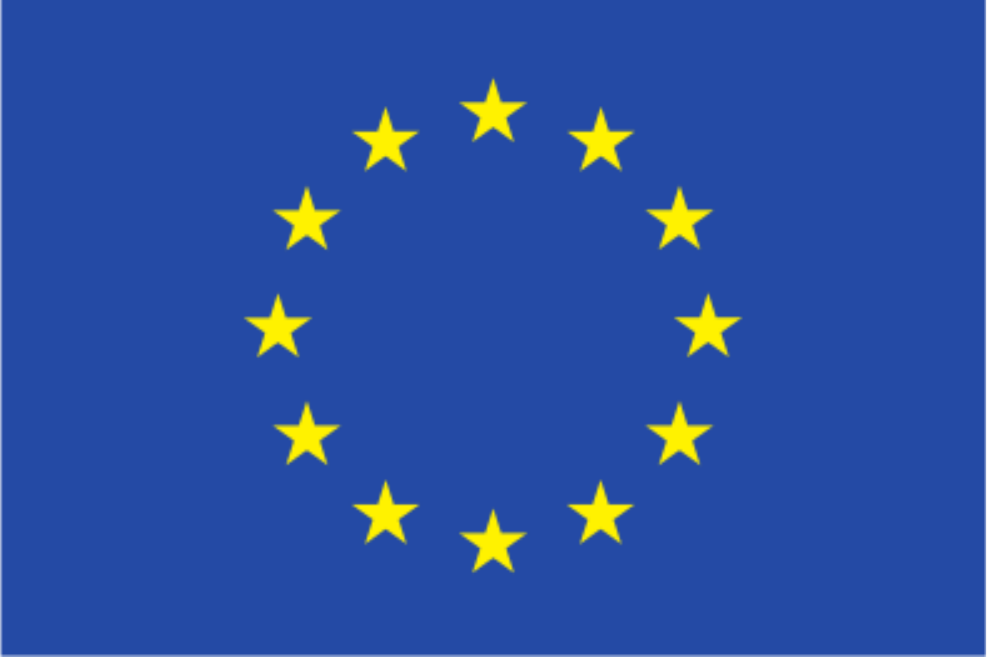}}

\title{Auto-active Verification of Floating-point Programs via Nonlinear Real Provers}

\titlerunning{Auto-active Verification of FP Programs via Nonlinear Real Provers}
\author{Junaid Rasheed\orcidID{0000-0002-2762-4097}
\and
Michal Konečný\orcidID{0000-0003-2374-9017}
\thanks{
This project has received funding from AdaCore Ltd and from \euflag{} the European Union’s Horizon 2020 research and innovation
programme under the Marie Skłodowska-Curie grant agreement No 731143.}
}
\authorrunning{J. Rasheed and M. Konečný}
\institute{Aston University, Birmingham, B4 7ET, UK
}
%

\maketitle              
\begin{abstract}
  We give a process for verifying numerical programs against their functional specifications.
  Our implementation is capable of automatically verifying programs against tight error bounds featuring common elementary functions.
  We demonstrate and evaluate our implementation on several examples, yielding the first fully verified SPARK implementations of the sine and square root functions.

  The process integrates existing tools using a series of transformations and derivations, building on the proving process in SPARK where Why3 produces Verification Conditions (VCs) and tools such as SMT solvers attempt to verify them.
  We add steps aimed specifically at VCs that contain inequalities with both floating-point operations and exact real functions.
  PropaFP is our open-source implementation of these steps.
  
  The steps include symbolic simplifications, deriving bounds via interval arithmetic, and safely replacing floating-point operations with exact operations, utilizing tools such as FPTaylor or Gappa to bound the compound rounding errors of expressions. 
  Finally, the VCs are passed to provers such as dReal, MetiTarski or LPPaver which attempt to complete the proof or suggest possible counter-examples.

  %
  
  


\end{abstract}

\section{Introduction}
\label{sec:introduction}
\paragraph{Context.}
Safety-critical software often includes numerical calculations.
Since most processors now contain a floating-point (FP) unit, these calculations often use FP arithmetic to utilise the speed and precision of FP units.

Those developing safety-critical programs need to provide guarantees that the program behaves in a precisely specified way.
This can be achieved via formal verification, i.e., proving that the program adheres to some specification.

\begin{figure}[tb]
  \begin{lstlisting}[caption=Sine function in Ada, language=ada, label={code:adaSineImpl}]
  function Taylor_Sin (X : Float) return Float is
    (X - ((X * X * X) / 6.0));
  \end{lstlisting}
\end{figure}

For example, consider the Ada function in Listing~\ref{code:adaSineImpl} that computes a Taylor approximation of the sine function.
We specify that this function gives a result very close to the exact sine function under some conditions:
\begin{equation}\label{eq:taylor-sine-informal-spec}
  X \in [-0.5, 0.5] \implies
  |\text{\lstinline{Taylor_Sin'Result}}-\sin(X)| \leq 0.001
\end{equation}
We would like a tool to automatically verify this specification or obtain a counter-example if it is not valid.
This is an example of auto-active verification \cite{leino2010auto-active}, i.e., automated proving of inline specifications such as post-conditions and loop invariants.

\begin{figure}[tb]
  \centering
  \includegraphics[width=0.8\hsize]{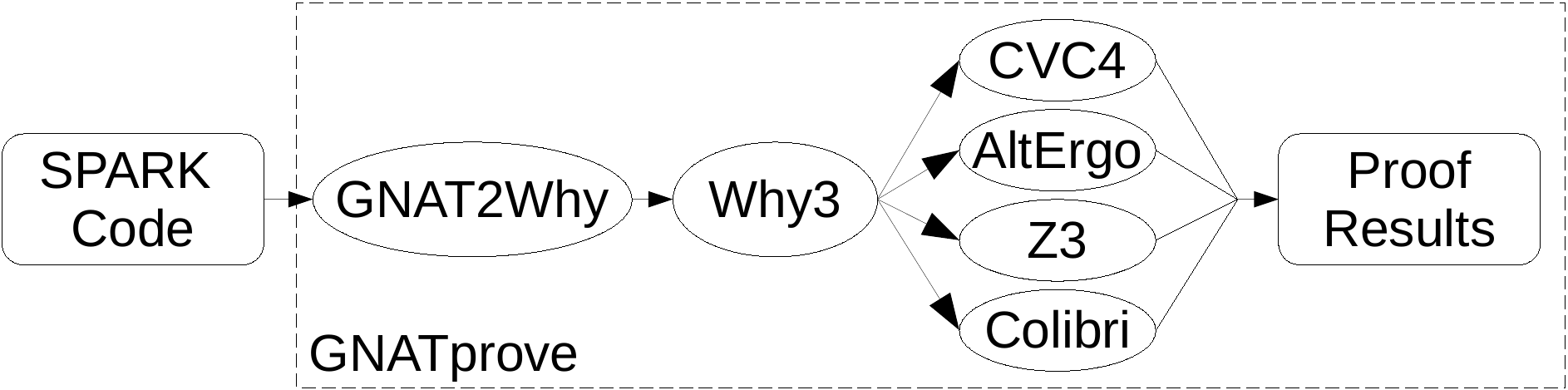}
  \caption{Overview of Automated Verification via GNATProve (adapted from \cite{fumex_automated_2017})}
  \label{fig:oldGnatprove}
\end{figure}

To this end, we deploy SPARK technology~\cite{hoang_spark_2015}, which represents the state-of-the-art in industry-standard formal software verification.
Specifically, we use SPARK Pro 22.1 which includes the GNAT Studio IDE and GNATprove.
GNATprove manages Why3 and a selection of bundled SMT solvers as shown in Fig.~\ref{fig:oldGnatprove}. 


As a language, SPARK is based on Ada with a focus on program verification.
GNATprove translates SPARK programs to WhyML programs using GNAT2Why.
Why3~\cite{bobot_why3_2011} then derives proof obligations in the form of verification conditions (VCs), which are formulas comprising traditional mathematical features such as numbers, numerical functions, and sets, and do not mention programming constructions such as loops and mutable variables.
The VCs imply that the program satisfies the given specification.
Finally, these VCs are sent to various SMT solvers which will attempt to decide them.
Why3 plays a key role in SPARK as well as other toolchains, effectively harnessing available solvers and provers for software verification.

\paragraph{Problem.}
With a SPARK version of the specification~(\ref{eq:taylor-sine-informal-spec}), the  toolchain automatically verifies absence of overflow in the \lstinline{Taylor_Sin} function.
This is not difficult since the input \lstinline{X} is restricted to the small domain $[-0.5,0.5]$.
However, the current SPARK toolchain and other frameworks we are aware of are unable to automatically verify that the result of \lstinline{Taylor_Sin(X)} is close to the exact $\sin(\text{\lstinline{X}})$.

Part of the problem is that the VCs feature a mixture of exact real and FP operations. 
For example, in the VCs derived from (\ref{eq:taylor-sine-informal-spec}), \lstinline{Taylor_Sin'Result} is replaced with
\[
  X \ominus ((X \otimes X \otimes X) \oslash 6.0);
\]
where $\ominus$, $\otimes$, and $\oslash$ are FP subtraction, multiplication. and division, respectively.
Although SPARK has some support for FP verification as described in \cite{fumex_automated_2017}, automatically verifying (\ref{eq:taylor-sine-informal-spec}) requires further work and that is what we set out to do here, building on \cite{fumex_automated_2017}.


\paragraph{Solution.}
To automatically verify functional specifications analogous to the one in equation~(\ref{eq:taylor-sine-informal-spec}), we have designed and implemented an extension of the SPARK proving process, called PropaFP.  The following steps are applied to quantifier-free VCs that contain real inequalities:
\begin{enumerate}
  \item Derive bounds for variables and simplify the VC.
  \item Safely replace FP operations with exact operations.
  \item Again simplify the VC.
  \item Attempt to decide the resulting VCs with provers for nonlinear real theorems.
\end{enumerate}
PolyPaver~\cite{duracz2014-polypaver} is a nonlinear real theorem prover that integrates with an earlier version of SPARK, but lacks the simplification steps and has much less powerful method of replacing FP operations.


\paragraph*{Paper outline.}

Section~\ref{sec:provingProcess} describes our process in detail, and
Section~\ref{sec:writing-specifications} analyses what constitutes an error bound provable by this process.
Sections~\ref{sec:verifyHeron} and \ref{sec:verifySin} illustrate the process on further examples, featuring a loop, domain reduction using integers, and calling non-trivial subprograms.
Section~\ref{sec:verifyHeron} presents an implementation of Heron's method for approximating the square-root of a number, while
Section~\ref{sec:verifySin} describes a verification of an adapted version of the sine function implementation from an AdaCore library used for safety-critical applications.
Section~\ref{sec:benchmarks} analyses the performance of the new proving process on the examples described in this paper and Section~\ref{sec:conclusion} concludes the paper.
%
%
%


\section{Our Proving Process}\label{sec:provingProcess}

When describing our proving process, we will illustrate its steps using the program \lstinline{Taylor_Sin} from Listing~\ref{code:adaSineImpl}.
Let us first consider its SPARK formal specification
shown in Listing \ref{code:sineSpec}.
To write more intuitive specifications, we use the Ada \lstinline{Big_Real} and \lstinline{Big_Integer} libraries to get rational arithmetic in specifications.
We have created axiomatic definitions of some exact functions, including sine which we call \lstinline{Real_Sin}.
These axiomatic functions have no implementation, only a specification which states that they behave like their analogous exact function.
The listings in this paper have shortened versions of some functions to aid readability.
Functions \lstinline{FC.To_Big_Real}, \lstinline{FLC.To_Big_Real}, and \lstinline{To_Real} respectively embed \lstinline{Floats}, \lstinline{Long_Floats} (doubles), and \lstinline{Integers} to \lstinline{Big_Reals}.
We have shortened these to \lstinline{Rf}, \lstinline{Rlf}, and \lstinline{Ri} respectively.
It should be understood that Why3 treats the \lstinline{Big_Real} type as reals, not rationals. 
The post-condition specifies a bound on the total error, i.e., the difference between this Taylor series approximation of sine and the exact sine of \lstinline{X}.

\begin{figure}[t]

\begin{lstlisting}[caption=SPARK formal specification of \lstinline{Taylor_Sin}, language=ada, label={code:sineSpec}]
function Taylor_Sin (X : Float) return Float with
  Pre => X >= -0.5 and X <= 0.5,
  Post => 
    abs(Real_Sin(Rf(X)) - Rf(Taylor_Sin'Result)) 
                          <= Ri(25889) / Ri(100000000); -- 0.00025889
\end{lstlisting}
\end{figure}

\begin{figure}[tb]
  \centering
  \includegraphics[width=0.8\hsize]{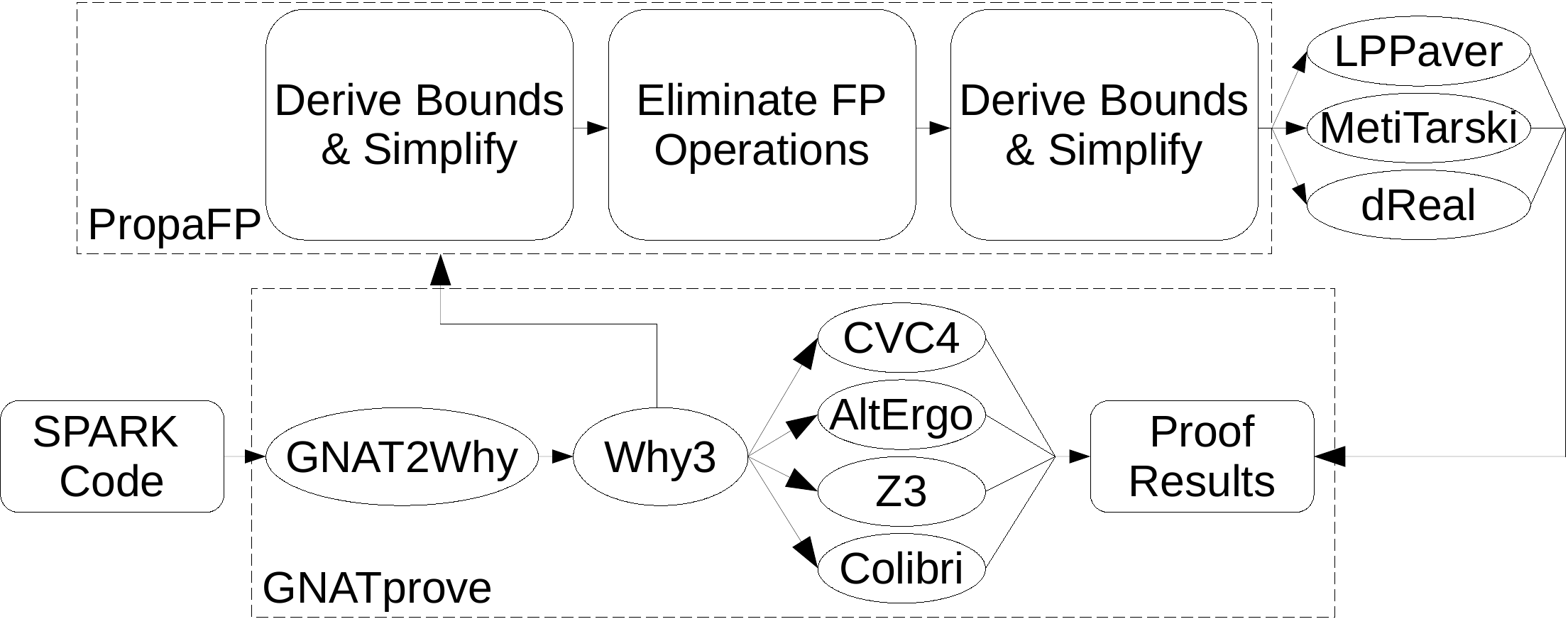}
  \caption{Overview of Automated Verification via GNATprove with PropaFP}
  \label{fig:gnatprove}
\end{figure}

\subsection{Generating and processing verification conditions}\label{sec:GeneratingVCs} 

As described in the introduction, we use GNATprove/Why3 to generate VCs.
In principle, we could use other programming and specification languages, as long as we can obtain VCs of a similar nature.

If a VC is not decided by the included SMT solvers, we use the Manual Proof feature in GNAT Studio to invoke PropaFP via a custom Why3 driver based on the driver for CVC4.  This driver first applies selected Why3 symbolic simplifications and saves the VC in SMT format.
As in this format the VC is a negation of the specification from which it was produced, we shall refer to it as `the negated VC' (NVC).
The VC \lstinline{contextAsConjunction $\implies$ goal} becomes the NVC \lstinline{contextAsConjunction $\wedge$ $\lnot$goal}.
During the process, we may weaken the conjunction of assertions by, for example, dropping assertions.
A model that satisfies the weakened NVC will not necessarily be a counter-example to the original VC or the original specification.
However, if the weakened NVC has no model, then both the original VC and the original specification are correct.

When parsing the SMT files, we 
ignore the definitions of basic arithmetic operations and transcendental functions.
Instead of using these definitions, we use each prover's built-in interpretations of such operations and functions.
In more detail, the parsing stage comprises the following steps:

\begin{itemize}
  \item Parse the SMT file as a list of symbolic expressions. Drop everything except assertions and variable and function type declarations.
  \item Attempt to parse all assertions in the conjunction. If an assertion contains any function not supported by the proving process, we drop the assertion.
  \item Functions are interpreted using their names.
  \begin{itemize}
    \item We rely on each prover's built-in understanding of the supported functions, currently $-,+,\times,\div,\exp, \log, \sin, \cos, \sqrt{\cdot}, \mathrm{mod}$.
    \item To increase the safety of this interpretation, we check the return type of some `ambiguous' functions.
    \begin{itemize} 
      \item For example, the output of a function named \lstinline{of_int} depends on the return type, i.e. If the return type of \lstinline{of_int(x)} is a single-precision float, parse this as \lstinline{Float32(x)}.
      \item For functions such as \lstinline{fp.add}, the return type is clear from the name of the function. Also, for these functions a type declaration is normally not included in the SMT file.
    \end{itemize} 
  \end{itemize} 
\end{itemize}
We use the operands to derive the type of the operation as follows:
\begin{itemize}
  \item FP operations are generic in the FP type, e.g., addition is encoded as \lstinline{fp.add param1 param2} for both single and double precision numbers. 
  \begin{itemize}
    \item FP literals are encoded in the SMT file as bits, making it trivial to determine their type.
    \item If the FP operation contains variables, use variable declarations to determine the type of each variable.
    \item If every literal and every variable in the operation has the same type, we can decide the type of the operation.
    \item If an operation has operands of differing types, we cannot determine the type of the FP operation, and drop the assertion. 
  \end{itemize}
\end{itemize}

\paragraph*{Dealing with $\pi$.}

Similar to \lstinline{Real_Sin}, we have created a function, \lstinline{Real_Pi}, which takes no input, and is specified with axioms to state that it behaves like $\pi$.
Why3 turns this into a function, \lstinline{real_pi}, with one \lstinline{void} input.
To help provers understand that this is the exact $\pi$, we substitute calls to the function \lstinline{real_pi} with $\pi$.

For \lstinline{Taylor_Sin}, the only VC that the SMT solvers included with GNAT Studio cannot solve is the post-condition VC.
The NVC for the post-condition is in Listing~\ref{vc-sine}.
It has been reformatted for better readability by, e.g., removing unneeded brackets, using circles for floating-point operations, and omitting irrelevant statements. 
The predicate \lstinline{isFiniteFloat(X)} is short for the inequalities \lstinline{MinFloat <= X, X <= MaxFloat}.

\begin{lstlisting}[float,caption=NVC corresponding to the post-condition from Listing~\ref{code:sineSpec}, label=vc-sine,basicstyle=\fontsize{7.5}{9pt}\tt]
-- assertions regarding axioms for sin and pi omitted
assert to_float(RNA, 1) $=$ 1.0
assert isFiniteFloat(x) 
assert (-0.5) $\le$ x $\wedge$ x $\le$ 0.5
assert isFiniteFloat(x$\odot$x) 
assert isFiniteFloat((x$\odot$x)$\odot$x) 
assert isFiniteFloat(x $\ominus$ (((x$\odot$x)$\odot$x)$\oslash$6.0)) 
assert
 $\neg$((
    sin(x) $+$ (-1$\cdot$(x $\ominus$ (((x$\odot$x)$\odot$x)$\oslash$6.0))) $\ge$ 0.0 
    $\implies$
    sin(x) $+$ (-1$\cdot$(x $\ominus$ (((x$\odot$x)$\odot$x)$\oslash$6.0))) $\le$ 25889/100000000
   )$\wedge$(
    $\neg$(sin(x) $+$ (-1$\cdot$(x $\ominus$ (((x$\odot$x)$\odot$x)$\oslash$6.0))) $\ge$ 0.0)
    $\implies$
    -1$\odot$(sin(x) $+$ (-1$\cdot$(x $\ominus$ (((x$\odot$x)$\odot$x)$\oslash$6.0)))) $\le$ 25889/100000000
   ))
\end{lstlisting}


\subsection{Simplifications and bounds derivation}\label{sec:simple-preprocessing}

As some of the tools used by PropaFP require bounds on all variables, we attempt to derive bounds from the assertions in the NVC.
First, we make the following symbolic simplifications to help derive better bounds:
\begin{itemize}
  \item Reduce vacuous propositions and obvious tautologies, such as:
  \begin{itemize}
    \item \lstinline{(NOT $\varphi$ OR true) AND ($\varphi$ OR false)} $\longrightarrow$ \lstinline{$\varphi$}
    \item \lstinline{$\varphi$ = $\varphi$} $\longrightarrow$ \lstinline{true}
  \end{itemize}
  \item Eliminate variables by substitution as follows: 
  \begin{itemize}
    \item Find variable-defining equations in the NVC, except  circular definitions.
    \item Pick a variable definition and make substitutions accordingly.
    \begin{itemize}
      \item E.g., pick \lstinline{i=i1+1}, and replace all occurrences of \lstinline{i} with \lstinline{i1+1}.
    \end{itemize} 
    \item If the variable has multiple definitions, pick the shortest one.
    \begin{itemize}
      \item E.g., if we have both \lstinline{x=1} and \lstinline {x=0+1}, all occurrences of x will be replaced with \lstinline{1}, including \lstinline{x=0+1} $\longrightarrow$ \lstinline{1=0+1}.
    \end{itemize}
  \end{itemize}
  \item Perform simple arithmetic simplifications, such as:
    \begin{itemize}
      \item \lstinline{$\varphi$ / 1} $\longrightarrow$ \lstinline{$\varphi$}
      \item \lstinline{0 + 1} $\longrightarrow$ \lstinline{1}
      \item \lstinline{MIN (e, e)} $\longrightarrow$ \lstinline{e}.
    \end{itemize}
  \item Repeat the above steps until no further simplification can be made.
\end{itemize}

Deriving bounds for variables proceeds as follows:
\begin{itemize}
  \item Identify inequalities which contain only a single variable on either side.
  \item Iteratively improve bounds by interval-evaluating the expressions given by these inequalities.
  \begin{itemize}
    \item Initially the bounds for each variable are $-\infty$ and $\infty$.
    \item For floating-point rounding $\mathit{rnd}(x)$, we overestimate the rounding error by the interval expression $x\cdot(1\pm\epsilon) \pm \zeta$ where $\epsilon$ is the machine epsilon, and $\zeta$ is the machine epsilon for denormalized numbers for the precision of the rounded operation.
  \end{itemize}
  \item These variables are assumed to be real unless they are declared integer. 
  \item For integer variables, 
  trim their bounds to nearest integers inside the interval.
\end{itemize} 

Next, use the derived bounds to potentially further simplify  the NVC:
\begin{itemize}
  \item Evaluate all formulas in the NVC using interval arithmetic.
  \item If an inequality is decided by this evaluation, replace it with \lstinline{True} or \lstinline{False}.
\end{itemize}

Finally, repeat the symbolic simplification steps, e.g., to remove any tautologies that have arisen in the interval evaluation.
Repeat deriving bounds, evaluations, and simplifications until we have no further improvement.

\paragraph{Similarities with Abstract Interpretation}

This process of deriving bounds can be thought of as a simple form of Abstract Interpretation (AI) over the interval domain, but instead of scanning program steps along paths in loops, we scan a set of mutually recursive variable definitions.
A similar iterative fixed-point calculation is used in both approaches.



The NVC arising from \lstinline{Taylor_Sin}, shown in Listing~\ref{vc-sine}, is already almost in its simplest form.  
The symbolic steps described in this section applied on this NVC only remove the assertions bounding \lstinline{X} and \lstinline{pi}, and replace \lstinline{pi} with $\pi$. 
The resulting NVC is outlined in Listing~\ref{vc-processed-sine}.

\begin{lstlisting}[float=t,caption=\lstinline{Taylor_Sin} NVC after simplification and bounds derivation, label=vc-processed-sine,basicstyle=\fontsize{7.5}{9pt}\tt]
Bounds on variables:
x (real) $\in$ [-0.5, 0.5] 

NVC:
assert to_float(RNA, 1) $=$ 1.0
-- The last assertion is unchanged from Listing~\ref{vc-sine} except turning $\geq$s into equivalent $\leq$s.
\end{lstlisting}

\subsection{Eliminating floating-point operations}\label{sec:elim:rounding}

VCs arising from floating-point programs are likely to contain floating-point operations.
As most provers for real inequalities do not natively support floating-point operations, we need to eliminate the floating-point operations before passing the NVCs to a numerical prover. 
We propose computing a bound on the size of the overall rounding errors in expressions using a tool specialised in this task,
replacing floating-point operations with exact operations, 
and compensating for the loss of rounding by adding/subtracting the computed error bound.
Note that this action weakens the NVCs. 
Recall that weakening is safe for proving correctness but may lead to incorrect counter-examples.

Currently, in our implementation we use FPTaylor \cite{solovyev2018-FPTaylor}, which supports most of the operations we need.
In principle, we can use any tool that gives reliable absolute bounds on the rounding error of our floating-point expressions, such as Gappa~\cite{10.1145/1644001.1644003} or Rosa~\cite{darulova2017-Rosa}.

There are expressions containing floating-point operations in the \lstinline{Taylor_Sin} NVC. 
The top-level expressions with FP operators are automatically passed to FPTaylor. 
Listing \ref{code:fptaylorHeron} shows an example of how the expressions are specified to FPTaylor.
The error bounds computed by FPTaylor for the \lstinline{Taylor_Sin} NVC expressions with floating-point operators are summarised in Table~\ref{table:rounding-errors}.

\begin{table}[ht]
\begin{center}
  \begin{tabular}{l@{\kern 1em}l}
    \toprule
    \fontsize{7}{10pt}\selectfont
    \texttt{rnd32(1.0)}                                                                   & 0 \\
    \fontsize{7}{10pt}\selectfont
    \texttt{sin(x) + (-1 * rnd32((x - rnd32((rnd32((rnd32((x * x)) * x)) / 6)))))}        & 1.769513e-8 \\
    \fontsize{7}{10pt}\selectfont
    \texttt{-1 * (sin(x) + (-1 * rnd32((x - rnd32((rnd32((rnd32((x * x)) * x)) / 6))))))} & 1.769513e-8 \\
    \bottomrule
  \end{tabular}
\end{center}
\caption{Error bounds computed by FPTaylor}\label{table:rounding-errors}
\end{table}

\begin{lstlisting}[float=t,caption=FPTaylor file to compute an error bound of the \lstinline{Taylor_Sin} VC,label={code:fptaylorHeron}, language=C]
Variables
  real x in [-0.5, 0.5];

Expressions
  sin(x) + (-1 * rnd32((x - rnd32((rnd32((rnd32((x*x))*x)) / 6)))));
// Computed absolute error bound: 1.769513e-8
\end{lstlisting}

We can now use these error bounds to safely replace FP operations with exact operations. 
Listing~\ref{vc-processed-sine-exact} shows the resulting NVC for \lstinline{Taylor_Sin}.

\begin{lstlisting}[float=t,caption={\lstinline{Taylor_Sin} NVC after removal of FP operations}, label=vc-processed-sine-exact,basicstyle=\fontsize{7.5}{9pt}\tt]
Bounds on variables:
x (real) $\in$ [-0.5, 0.5] 

NVC: 
assert 1 $\pm$ $\texttt{\uline{0.0}}$ $=$ 1.0 
assert
 $\neg$((
  0.0 $\le$ (sin(x) $+$ (-1$\cdot$(x $-$ ((x$\cdot$x)$\cdot$x$/$6.0))) $+$ $\uline{1.769513\texttt{e}^{-8}}$)
  $\implies$
  (sin(x) $+$ (-1$\cdot$(x $-$ ((x$\cdot$x)$\cdot$x$/$6.0))) $+$ $\uline{1.769513\texttt{e}^{-8}}$) $\le$ (25889/100000000) 
 )$\wedge$(
  $\neg$ (0.0 $\le$ (sin(x) $+$ (-1$\cdot$(x $-$ ((x$\cdot$x)$\cdot$x$/$6.0))) $-$ $\uline{1.769513\texttt{e}^{-8}}$)) 
  $\implies$
  (-1$\cdot$(sin(x) $+$ (-1$\cdot$(x $-$ ((x$\cdot$x)$\cdot$x$/$6.0))) $+$ $\uline{1.769513\texttt{e}^{-8}}$)) $\le$ (25889/100000000)
 ))
\end{lstlisting}

There may be statements which can be further simplified thanks to the elimination of FP operations.
For example, in Listing~\ref{vc-processed-sine-exact}, we have the trivial tautology $1 \pm 0.0 = 1.0$.
To capitalise on such occurrences, we 
could once again interval-evaluate each statement in the NVC.
Instead, we invoke the steps from Section~\ref{sec:simple-preprocessing} again, which not only include interval evaluation, but also make any consequent simplifications.
In Table \ref{table:benchmarks} this NVC is referred to as Taylor\_Sin.

We now have derived bounds for variables and a weakened and simplified NVC with no FP operations, ready for provers.  We will call this the `simplified exact NVC'.

\begin{lstlisting}[float=t,caption={\lstinline{Taylor_Sin} simplified exact NVC, ready for provers}, label=vc-processed-sine-exact-s,basicstyle=\fontsize{7.5}{9pt}\tt]
Bounds on variables:
x (real) $\in$ [-0.5, 0.5] 

NVC: 
-- The last assertion is the same as in Listing~\ref{vc-processed-sine-exact}
\end{lstlisting}

\paragraph{Alternative Methods to Deal with Floating-Point Operations}
Why3 includes a formalization of the FP IEEE-754 standard \cite{noauthor_ieee_2008}.
For SMT solvers that natively support FP operations, this formalization is mapped to the SMT-LIB FP theory, and
for SMT solvers that do not support FP operations, an axiomatization of the formalization is given \cite{fumex_automated_2017}.
This approach is not powerful enough to verify our \lstinline{Taylor_Sin} NVC as well as examples we discuss later.

\section{Deriving Provable Error Bounds}\label{sec:writing-specifications}

The specification in Listing~\ref{code:sineSpec} bounds the difference between \lstinline{Taylor_Sin(X)} and the exact sine function.
Such a bound can be broken down as follows:
\begin{itemize}
  \item The \textbf{subprogram specification error}, i.e. the error inherited from the specification of any subprograms that the implementation relies on.
  \begin{itemize}
    \item If an implementation relies on some subprogram, the specification, not the implementation, of that subprogram would be used in the Why3 VC.
    \item For \lstinline{Taylor_Sin} this component is $0$ as it does not call any subprograms.
  \end{itemize} 
  \item The \textbf{maximum model error}, ie the maximum difference between the \emph{model} used in the computation and the exact \emph{intended} result.
  \begin{itemize}
    \item For \lstinline{Taylor_Sin} this is the difference between the degree 3 Taylor polynomial for the sine function and the sine function.
  \end{itemize}
  \item The \textbf{maximum rounding error}, i.e. the maximum difference between the \emph{exact model} and the \emph{rounded model} computed with FP arithmetic.
  \item A \textbf{rounding analysis cushion} arising when eliminating floating-point operations.  This is the difference between the \emph{actual maximum} rounding error and the \emph{bound} on the rounding error calculated by a tool such as FPTaylor as well as over-approximations made when deriving bounds for variables. 
  \begin{itemize}
    \item The derived bounds are imperfect due to the accuracy loss of interval arithmetic as well as the over-approximation of floating-point operations.
    \item Imperfect bounds inflate the computed rounding error bound, as more values have to be considered.
  \end{itemize} 
  \item A \textbf{proving cushion} is added so that the specification can be decided by the approximation methods in the provers.
  Without this \emph{cushion}, the provers could not decide the given specification within certain bounds on resources, such as a timeout.
\end{itemize}

To justify our specification in Listing~\ref{code:sineSpec}, we estimated the values of all five components. 
Our estimates can be seen in Table \ref{table:taylor-sine-components}.
The \textbf{maximum model error} and the \textbf{maximum rounding error} were calculated using the Monte-Carlo method.
We ran a simulation comparing the Taylor series approximation of degree 3 of the sine function and an exact sine function.
This simulation was ran for one million with pseudo-random inputs, giving us an approximate model error.
To estimate the maximum rounding error,
we compared a single precision and a quadruple precision floating-point implementation of the model for one hundred million pseudo-random inputs.
(Floating-point operations are much faster than exact operations.)
We estimated \textbf{rounding analysis cushion} by the difference between the \textbf{rounding errror} and the bound given by FPTaylor ($\sim 1.77E{-}8$).
Note that the actual \textbf{rounding analysis cushion} may be larger due to over approximations made when deriving bounds.

\begin{table}[t]
  \begin{center}
    \begin{tabular}{l@{\kern 1em}l@{\kern 1em}l}
      \toprule
      & single precision & double precision \\
      \cmidrule(r){2-2}
      \cmidrule{3-3}
      \textbf{Subprogram Specification Error} & $0$  & $0$\\
      \textbf{Maximum Model Error}            & $\sim 2.59E{-}4$ & $\sim 2.59E{-}4$ \\
      \textbf{Maximum Rounding Error}         & $\sim 1.61E{-}8$ & $\sim 2.89E{-}17$ \\
      \textbf{Rounding Analysis Cushion}         & $\sim 1.57E{-}9$ & $\sim 4.04E{-}18$\\ 
      \textbf{Proving Cushion}       & $\sim 2.11E{-}9$ & $\sim 1.80E{-}9$\\
      \bottomrule
    \end{tabular}
  \end{center}
  \caption{Error bound components for \lstinline{Taylor_Sin}}\label{table:taylor-sine-components}
\end{table}

The sum of the \textbf{maximum model error}, the \textbf{maximum rounding error}, and the \textbf{rounding analysis cushion} 
is around 0.0002588878950.
Raising the specification bound to 0.00025889 enables provers LPPaver and dReal to verify the specification, using a \textbf{proving cushion} of around $2.11E{-}9$.


In this case, most of the error in the program comes from the \textbf{maximum model error}.
If we increased the number of Taylor terms, the \textbf{maximum model error} would become smaller and the \textbf{maximum rounding error} would become larger.
Increasing the input domain would make both the \textbf{maximum model error} and the \textbf{maximum rounding error} larger. 

Increasing the precision of the floating-point numbers used is a simple way to reduce both the maximum \textbf{rounding error} and the \textbf{rounding analysis cushion}.
Table~\ref{table:taylor-sine-components} on the right shows estimates for the components in a double-precision version of \lstinline{Taylor_Sin}.
The simplified exact NVC resulting from this example is referred to as Taylor\_Sin\_Double in Table~\ref{table:benchmarks}.


To see how the \textbf{subprogram specification error} affects provable error bounds, consider function \lstinline{SinSin}
given in Listings~\ref{code:sinSinBody} and \ref{code:sinSinSpec}.

\begin{figure}[ht]
\vspace{-2ex}
\begin{lstlisting}[caption=SinSin function definition in SPARK, language=ada, label={code:sinSinBody}]
procedure Taylor_Sin_P (X : Float; R : out Float) is
begin
    R := X - ((X * X * X) / 6.0);
end Taylor_Sin_P;

function SinSin (X : Float) return Float is
    OneSin, TwoSin : Float;
begin
    Taylor_Sin_P(X, OneSin);
    Taylor_Sin_P(OneSin, TwoSin);
    return TwoSin;
end SinSin;
\end{lstlisting}

\begin{lstlisting}[caption=SinSin function specification in SPARK, language=ada, label={code:sinSinSpec}]
procedure Taylor_Sin_P (X : Float; R : out Float) with
  Pre => X >= -0.5 and X <= 0.5,
  Post =>
    Rf(R) >= Ri(-48) / Ri(100) and -- Helps verification of calling functions
    Rf(R) <= Ri(48) / Ri(100) and       
    abs(Real_Sin(Rf(X)) - Rf(R)) <= Ri(25889) / Ri(100000000);

function SinSin ( X : Float) return Float with
  Pre => X >= -0.5 and X <= 0.5,
  Post =>
    abs(Real_Sin(Real_Sin(Rf(X))) - Rf(SinSin'Result))
     <= Ri(51778) / Ri(100000000);
\end{lstlisting}
\end{figure}

\lstinline{Taylor_Sin_P} is the procedure version of the \lstinline{Taylor_Sin} function. 
Our implementation currently does not support function calls, but it does support procedure calls. (This limitation is not conceptually significant.)
The specification for \lstinline{Taylor_Sin_P} has two additional inequalities, bounding the output value \lstinline{R} to allow us to derive tight bounds for \lstinline{R} when proving VCs involving calls of this procedure.
Verifying this procedure in GNATprove gives one NVC for our proving process, corresponding to the final post-condition.
This NVC is referred to as and Taylor\_Sin\_P in Table~\ref{table:benchmarks}.
The exact NVC is in folder \href{https://github.com/rasheedja/PropaFP/tree/master/examples/taylor_sine}{\texttt{examples/taylor\_sine}} in \cite{noauthor_rasheedjapropafp_nodate}.







Function \lstinline{SinSin} calls \lstinline{Taylor_Sin_P} with the parameter \lstinline{X}, storing the result in variable \lstinline{OneSin}.
\lstinline{Taylor_Sin_P} is then called again with the parameter \lstinline{OneSin}, storing the result in \lstinline{TwoSin}, which is then returned.
The post-condition for the \lstinline{SinSin} function specifies the difference between our \lstinline{SinSin} implementation and calling the exact sine function on \lstinline{X} twice.
The VC resulting from this post-condition is referred to as SinSin in Table \ref{table:benchmarks}.



Since the steps of \lstinline{SinSin} involve only subprogram calls, there is no \textbf{model error} or \textbf{rounding error}, and thus no \textbf{rounding analysis cushion}.
As the value of \lstinline{SinSin} comes from \lstinline{Taylor_Sin_P} applied twice, and the derivative of $\sin$ has the maximum value $1$, the \textbf{subprogram specification error} is a little below $0.00025889 + 0.00025889 = 0.00051778$.
Experimenting with different bounds, we estimate the 
LPPaver \textbf{proving cushion} is around $10^{-13}$.


There is a delicate trade-off between the five components that a programmer would need to manage by a careful choice of the model used, floating-point arithmetic tricks, and proof tools used to obtain a specification for a program that is both accurate and does not require large cushions or specification errors.
It is not our goal to make this type of optimisation for the example programs, rather we have calculated these values to help improve the understanding of how difficult it is to estimate them in practice.
In simple cases, it would be sufficient to tighten and loosen the `bound' in the specification until the proving process succeeds and fails, respectively.

%

\section{Verification of Heron's Method for Approximating the Square Root Function}\label{sec:verifyHeron}
We used PropaFP to verify an implementation of Heron's method.
This is an interesting case study because it requires the use of loops and loop invariants.

\begin{figure*}[ht]
\vspace{-2ex}
\begin{lstlisting}[caption=Heron's Method Specification, language=ada, label={code:heronSpec},basicstyle=\fontsize{7.5}{9pt}\tt]
function Certified_Heron (X : Float; N : Integer) Return Float with
  Pre => X >= 0.5 and X <= 2.0 and N >= 1 and N <= 5,
  Post =>
    abs(Real_Square_Root(Rf(X)) - Rf(Certified_Heron'Result))
      <= (Ri(1) / (Ri(2 ** (2 ** N)))) -- $\smash{\nicefrac{1}{2^{2^{\mathtt{N}}}}}$ model error
             + Ri(3*N)*(Ri(1)/Ri(8388608)); --  $3\cdot\mathtt{N}\cdot \varepsilon$, rounding error bound
\end{lstlisting}

\begin{lstlisting}[caption=Heron's Method Implementation, language=ada, label={code:heronImpl},basicstyle=\fontsize{7.5}{9pt}\tt]
function Certified_Heron (X : Float; N : Integer) return Float is
  Y : Float := 1.0;
begin
  for i in 1 .. N loop
    Y := (Y + X/Y) / 2.0;

    pragma Loop_Invariant (Y >= 0.7);
    pragma Loop_Invariant (Y <= 1.8);
    pragma Loop_Invariant
      (abs (Real_Square_Root (Rf(X)) - Rf(Y))
        <= (Ri(1) / (Ri(2 ** (2 ** N)))) -- $\smash{\nicefrac{1}{2^{2^{\mathtt{i}}}}}$
              + Ri(3*i)*(Ri(1)/Ri(8388608))); -- $3\cdot\mathtt{i}\cdot \varepsilon$
  end loop;
  return Y;
end Certified_Heron;
\end{lstlisting}
\vspace{-2ex}
\end{figure*}

In Listing~\ref{code:heronSpec}, the term $3\cdot\mathtt{N}\cdot \varepsilon$ is a bound for the compound rounding error guessed by counting the number of operations.
Note that five iterations are more than enough to get a good approximation of the square root function for X in the range $[0.5, 2]$.


The implementation in Listing \ref{code:heronImpl} contains the loop invariants.
The bounds on Y here help generate easier VCs for the loop iterations and post-loop behaviour.
The main loop invariant is very similar to the post-condition in the specification, except substituting \lstinline{i} for \lstinline{N}, essentially specifying the difference between the exact square root and Heron's method for each iteration of the loop.

Why3 produces 74 NVCs from our implementation of Heron's method.
72 of these NVCs are either trivial or verified by SMT solvers. 
PropaFP is required for 2 NVCs that come from the main loop invariant. 
One NVC specifies that the loop invariant holds in the initial iteration of the loop, where \lstinline{i} is equal to 1.
Another VC specifies that the loop invariant is preserved from one iteration to the next, where \lstinline{i} ranges from 1 and \lstinline{N}.
We refer to these NVCs as  as Heron\_Init and Heron\_Pres in Table \ref{table:benchmarks}.
Note that the third NVC derived from the invariant, i.e., that the invariant on the last iteration implies the postcondition is trivial here.
The corresponding simplified exact NVCs can be found in folder \href{https://github.com/rasheedja/PropaFP/tree/master/examples/heron/txt}{\texttt{examples/heron}} in \cite{noauthor_rasheedjapropafp_nodate}.

\section{Verifying AdaCore's Sine Implementation}\label{sec:verifySin}

With the help of PropaFP, we developed a verified version of an Ada sine implementation written by AdaCore for their high-integrity math library\footnote{We obtained the original code from file \texttt{src/ada/hie/s-libsin.adb} in archive \texttt{gnat-2021-20210519-19A70-src.tar.gz} downloaded from ``More packages, platforms, versions and sources'' at \url{https://www.adacore.com/download}.}.
First, we translated the program from Ada to SPARK, which entailed removing the use of generic types and other SPARK-violating code.

The code consists of several dependent subprograms.  
There are functions for computing $\sin(x)$ and $\cos(x)$ for $x$ close to $0$ and functions that extend the domain to $x\in [-802,802]$ by translating $x$ into one of the four basic quadrants near $0$.
There is also a loop that extends the domain further.
We have focused on the code for $x \in [-802,802]$ and postponed the verification of the loop.

We have translated functions into procedures since PropaFP currently does not support function calls.
Next, we discuss all six procedures that we needed to specify and verify.


\subsection{Multiply\_Add}

\begin{figure*}[t]
\vspace{-2ex}
\begin{lstlisting}[caption=Multiply\_Add Implementation, language=ada, label={code:HIEMultiplyAddImpl},basicstyle=\fontsize{7.5}{9pt}\tt]
procedure Multiply_Add (X, Y, Z : Float; Result : out Float) is
begin
    Result := (X * Y + Z);
end Multiply_Add;
\end{lstlisting}
  
\begin{lstlisting}[caption=Multiply\_Add Specification, language=ada, label={code:HIEMultiplyAddSpec},basicstyle=\fontsize{7.5}{9pt}\tt]
  procedure Multiply_Add (X, Y, Z : Float; Result : out Float) with
  Pre => 
    (-3.0 <= X and X <= 3.0) and
    (-3.0 <= Y and Y <= 3.0) and
    (-3.0 <= Z and Z <= 3.0),
  Post =>
    (-12.0 <= Result and Result <= 12.0) and
    Result = X * Y + Z;
\end{lstlisting}
\vspace{-2ex}
\end{figure*}

The specification in Listing \ref{code:HIEMultiplyAddSpec} restricts the ranges of the input and output to rule out overflows.
We used very small bounds based on how the function is used locally by the other procedures.


\subsection{My\_Machine\_Rounding}

\begin{figure*}[tb]
\vspace{-2ex}
\begin{lstlisting}[caption=My\_Machine\_Rounding Implementation, language=ada, label={code:HIERoundingImpl},basicstyle=\fontsize{7.5}{9pt}\tt]
procedure My_Machine_Rounding (X : Float; Y : out Integer) is
begin
    Y := Integer(X); -- rounding to nearest
end My_Machine_Rounding;
\end{lstlisting}
  
\begin{lstlisting}[caption=My\_Machine\_Rounding Specification, language=ada, label={code:HIERoundingSpec},basicstyle=\fontsize{7.5}{9pt}\tt]
procedure My_Machine_Rounding (X : Float; Y : out Integer) with
  Pre =>
    (0.0 <= X and X <= 511.0),
  Post =>
    (0 <= Y and Y <= 511) and
    Rf(X) - Ri(Y) >= Ri(-500000001) / Ri(1000000000) and -- -0.500000001
    Rf(X) - Ri(Y) <= Ri(500000001) / Ri(1000000000);     --  0.500000001
\end{lstlisting}
\vspace{-2ex}
\end{figure*}

This is a custom procedure that is used to round a floating-point number to the nearest integer.
In the original version of this code, this was done using the SPARK-violating Ada function, \lstinline{Float'Machine_Rounding}.

Again, we specify the ranges of the variables based on the local use of this procedure, to make it easier for our provers to verify the resulting VCs.

The other post-conditions state that the difference between \lstinline{X} and \lstinline{Y} (which is \lstinline{X} rounded to the nearest integer) is, at most, 0.500000001.
We chose this number to avoid any ``touching'' VCs (such as $x > 0 \implies x > 0$), which solvers using interval methods usually cannot prove.  
While SMT solvers can usually verify simple touching VCs, here they fail, probably due to the rounding function.

The NVCs resulting from the last two post-conditions are referred to as My\_Machine\_Rounding${}_\ge$ and My\_Machine\_Rounding${}_\le$ in Table \ref{table:benchmarks}.

\subsection{Reduce\_Half\_Pi}

This procedure takes some input value, \lstinline{X}, and subtracts a multiple of $\frac{\pi}{2}$ to translate it into the interval $[-0.26 \otimes \textit{floatingPointPi}, 0.26 \otimes \textit{floatingPointPi}]$.

\begin{figure*}[ht]
\vspace{-2ex}
\begin{lstlisting}[caption=Reduce\_Half\_Pi Implementation, language=ada, label={code:HIEReduceImpl},basicstyle=\fontsize{7.5}{9pt}\tt]
procedure Reduce_Half_Pi (X : in out Float; Q : out Quadrant; R : out Integer) 
is
  K      : constant       := Pi / 2.0;
  --  Bits\_N : constant       := 9;
  --  Bits\_C : constant       := Float'Machine\_Mantissa - Bits\_N;
  C1     : constant Float := 1.57073974609375; 
                              -- Float'Leading\_Part (K, Bits\_C);
  C2     : constant Float := 0.0000565797090530395508; 
                              -- Float'Leading\_Part (K - C1, Bits\_C);
  C3     : constant Float := 0.000000000992088189377682284; 
                              -- Float'Leading\_Part (K - C1 - C2, Bits\_C);
  C4     : constant Float := K - C1 - C2 - C3;
  N      : Float := (X / K); 
begin
  My_Machine_Rounding(N, R);   -- R is returned for use in the specification
  
  X := (((X - Float(R)*C1) - Float(R)*C2) - Float(R)*C3) - Float(R)*C4;
  -- The above is roughly equivalent to X := (X - Float(R)*K);
  Q := R mod 4;
end Reduce_Half_Pi;
\end{lstlisting}
  
\begin{lstlisting}[caption=Reduce\_Half\_Pi Specification, language=ada, label={code:HIEReduceSpec},basicstyle=\fontsize{7.5}{9pt}\tt]
subtype Quadrant is Integer range 0 .. 3;

Max_Red_Trig_Arg : constant := 0.26 * Ada.Numerics.Pi;
Half_Pi          : constant := Ada.Numerics.Pi / 2.0;

procedure Reduce_Half_Pi (X : in out Float; Q : out Quadrant; R : out Integer)
  with Pre =>  X >= 0.0 and X <= 802.0,
  Post => 
    R >= 0 and R <= 511 and
    Rf(X'Old / (Pi/2.0)) - Ri(R) >= Ri(-500000001)/Ri(1000000000) and
    Rf(X'Old / (Pi/2.0)) - Ri(R) <= Ri(500000001)/Ri(1000000000) and
    Q = R mod 4 and 
    X >= -Max_Red_Trig_Arg and X <= Max_Red_Trig_Arg and
    (Rf(X) - (Rf(X'Old) - (Ri(R)*Real_Pi/Rf(2.0)))) >= Ri(-18)/Ri(100000) 
    and
    (Rf(X) - (Rf(X'Old) - (Ri(R)*Real_Pi/Rf(2.0)))) <= Ri(18)/Ri(100000);    
\end{lstlisting}
\vspace{-2ex}
\end{figure*}

The implementation, seen in Listing \ref{code:HIEReduceImpl}, has some significant differences to the original implementation.
First, we limited this procedure to \lstinline{X} within $[0, 802]$ and removed a loop that catered for larger values, as mentioned earlier.
Also, we inlined calls to the SPARK-violating function \lstinline{Float'Leading_Part}, which removes a specified number of bits from a floating-point number.
This function was used to define the variables \lstinline{C1}, \lstinline{C2}, and \lstinline{C3}, in effect, giving a higher precision version of $\pi / 2$ using single-precision floating-point variables.

The specification in Listing \ref{code:HIEReduceSpec} uses the new out parameter \lstinline{R}, which was just a local variable in the original implementation.
\lstinline{R} holds the integer multiple of $\frac{\pi}{2}$ used to shift the input value close to $0$.
The final two post-conditions bound the difference between the computed new value of \lstinline{X} and the ideal model result.

Our proving process is needed for the NVCs derived from the last four post-conditions in Listing \ref{code:HIEReduceSpec}.
These four NVCs derived are referred to in Table \ref{table:benchmarks} as Reduce\_Half\_Pi\_X\{${}_\ge$,${}_\le$\} and Reduce\_Half\_Pi\{${}_\ge$,${}_\le$\}, respectively.

\subsection{Approx\_Sin and Approx\_Cos}

\begin{figure*}[p]
\vspace{-2ex}
\begin{lstlisting}[caption=Approx\_Sin and Approx\_Cos Implementation, language=ada, label={code:HIEApproxImpl},basicstyle=\fontsize{7.5}{9pt}\tt]
procedure Approx_Sin  (X : Float; Result : out Float) is
  Sqrt_Epsilon_LF : constant Long_Float :=
    Sqrt_2 ** (1 - Long_Float'Machine_Mantissa); 

  G  : constant Float := X * X;
  
  -- Horner Scheme
  H0 : constant Float := (-0.19501_81843E-3);
  H1 : Float;
  H2 : Float;
begin
  Multiply_Add(H0, G, (0.83320_16396E-2), H1);
  Multiply_Add(H1, G, (-0.16666_65022), H2);
  if abs X <= Float(Long_Float (Sqrt_Epsilon_LF)) then
     Result := X;
  else
     Result := (X * (H2 * G) + X);
  end if;
end Approx_Sin;

procedure Approx_Cos (X : Float; Result : out Float) is
  G  : constant Float := X * X;
  
  -- Horner Scheme
  H0 : constant Float := (0.24372_67909E-4);	
  H1 : Float;
  H2 : Float;
  H3 : Float;
  H4 : Float;
begin
  Multiply_Add(H0, G, (-0.13888_52915E-2), H1);
  Multiply_Add(H1, G, (0.41666_61323E-1), H2);
  Multiply_Add(H2, G, (-0.49999_99957), H3);
  Multiply_Add(H3, G, (0.99999_99999), H4);
  Result := H4;
end Approx_Cos;
\end{lstlisting}
  
\begin{lstlisting}[caption=Approx\_Sin and Approx\_Cos Specification, language=ada, label={code:HIEApproxSpec},basicstyle=\fontsize{7.5}{9pt}\tt]
Max_Red_Trig_Arg : constant := 0.26 * Ada.Numerics.Pi;
Sqrt_2 : constant := 1.41421_35623_73095_04880_16887_24209_69807_85696;

procedure Approx_Sin  (X : Float; Result : out Float) with
  Pre  =>
    X >= -Max_Red_Trig_Arg and X <= Max_Red_Trig_Arg,
  Post =>
    Result >= -1.0 and Result <= 1.0 and
    (Rf(Result) - Real_Sin(Rf(X))) >= Ri(-58) / Ri(1000000000) and
    (Rf(Result) - Real_Sin(Rf(X))) <= Ri(58) / Ri(1000000000);

procedure Approx_Cos (X : Float; Result : out Float) with
  Pre  =>
    X >= -Max_Red_Trig_Arg and X <= Max_Red_Trig_Arg,
  Post =>
    Result >= -1.0 and Result <= 1.0 and
    (Rf(Result) - Real_Cos(Rf(X))) >= Ri(-14) / Ri(100000000) and
    (Rf(Result) - Real_Cos(Rf(X))) <= Ri(14) / Ri(100000000);
\end{lstlisting}
\vspace{-2ex}
\end{figure*}

Approx\_Sin and Approx\_Cos in Listing \ref{code:HIEApproxImpl} compute Taylor series approximations of sine and cosine, respectively, using the Horner scheme.
%
In the original AdaCore implementation, variable \lstinline{X} has a generic type, but 
we have fixed the type to \lstinline{Float}.
The original implementation uses arrays and loops to adapt the order of the Taylor series to the precision of the float type.
Since we have fixed the type of X, we perform these computations directly without arrays and loops.

The specifications in Listing \ref{code:HIEApproxSpec} are quite simple.
The pre-conditions restrict the value of \lstinline{X} to be within the interval $[-0.26 \otimes \textit{floatingPointPi}, 0.26 \otimes \textit{floatingPointPi}]$.
The first two post-conditions in both procedures restrict the \lstinline{Result} to be within the interval $[-1,1]$.
The last two post-conditions in both procedures specify the difference between the exact version of Sine/Cosine and Approx\_Sin/Approx\_Cos.


The NVCs corresponding to the last two postconditions in both procedures
are called Approx\_Sin\{${}_\ge$,${}_\le$\} and Approx\_Cos\{${}_\ge$,${}_\le$\} in Table~\ref{table:benchmarks}.

\subsection{Sin}

\begin{figure*}[t]
\vspace{-2ex}
\begin{lstlisting}[caption=Sin Implementation, language=ada, label={code:HIESinImpl},basicstyle=\fontsize{7.5}{9pt}\tt]
procedure Sin (X : Float; FinalResult : out Float) is
  Y      : Float := (if X < 0.0 then -X else X);
  Q      : Quadrant;
  R      : Integer;
  Result : Float;
begin
  Reduce_Half_Pi (Y, Q, R);
  
  if Q = 0 or Q = 2 then
     Approx_Sin (Y, Result);
  else -- Q = 1 or Q = 3
     Approx_Cos (Y, Result);
  end if;
  
  if X < 0.0 then
     FinalResult := (-1.0) * (if Q >= 2 then -Result else Result);
  else
     FinalResult := (1.0)  * (if Q >= 2 then -Result else Result);
  end if;
end Sin;  

\end{lstlisting}
  
\begin{lstlisting}[caption=Sin Specification, language=ada, label={code:HIESinSpec},basicstyle=\fontsize{7.5}{9pt}\tt]
procedure Sin (X : Float; FinalResult : out Float)
  with Pre => 
    X >= -802.0 and X <= 802.0,
  Post =>
    (Rf(FinalResult) - Real_Sin(Rf(X))) >= Ri(-19) / Ri(100000) and
    (Rf(FinalResult) - Real_Sin(Rf(X))) <= Ri(19) / Ri(100000);
\end{lstlisting}
\vspace{-2ex}
\end{figure*}

Finally, procedure Sin in Listing~\ref{code:HIESinImpl} approximates the sine function for inputs from $[-802,802]$.
Compared to the original function, 
we have replaced uses of the SPARK-violating function \lstinline{Float'Copy_Sign} with code that has the same effect.

Our proving process is needed to verify NVCs arising from the final two post-conditions in Listing \ref{code:HIESinSpec}, and these are referred to as Sin${}_{\{\ge,\le\}}$ in Table \ref{table:benchmarks}.

\subsection{Generated Why3 NVCs}
In total, Why3 derives 158 NVCs from the six procedures we have described.
SMT solvers can verify 146 NVCs.
The 12 remaining NVCs can be verified using our proving process.
This is broken down by procedure in Table~\ref{table:hieVCs}. 

\begin{table}[t]
  \caption{Why3 NVCs Generated for each Procedure from our Modified AdaCore Sine Implementation}
  \label{table:hieVCs}
  \centering
  \begin{tabular}{l@{\kern1em}r@{\kern1em}r@{\kern1em}r}
  \toprule
  Procedure             & Generated NVCs & Trivial/SMT & Proving 
  Process \\
  \midrule
  Multiply\_Add         & 4                       & 4                             & 0                                           \\
  My\_Machine\_Rounding & 16                       & 14                           & 2                                           \\
  Reduce\_Half\_Pi      & 44                      & 40                            & 4                                           \\
  Approx\_Sin           & 33                      & 31                            & 2                                           \\
  Approx\_Cos           & 41                      & 39                            & 2                                           \\
  Sin                   & 20                      & 18                             & 2                                           \\
  \bottomrule
  \end{tabular}
\end{table}

We discuss only a few of the more interesting NVCs here.
All NVCs can be found in folder \href{https://github.com/rasheedja/PropaFP/tree/master/examples/hie_sine/txt}{\texttt{examples/hie\_sine}} in \cite{noauthor_rasheedjapropafp_nodate}.

\begin{lstlisting}[caption=Selected Reduce\_Half\_Pi Simplified Exact NVCs, label={code:HIEReduceVCs},basicstyle=\fontsize{7.5}{9pt}\tt,float=tp]
Reduce_Half_Pi_X_${}_\le$

Bounds on variables:
r1 (int) $\in$ [0, 511]
x (real) $\in$ [0, 802]

assert
  -500000001 / 1000000000 <=
    (((x / (13176795/8388608)) - r1) + (542101/2500000000000000000000000))

assert
  (((x / (13176795/8388608)) - r1) - (542101/2500000000000000000000000)) <=
    500000001 / 1000000000

assert
  $\neg$(
    ((((x - (r1 * (25735/16384))) - (r1 * (3797/67108864))) 
       - (r1 * (17453/17592186044416)))
     - (r1 * (12727493/2361183241434822606848))) + (1765573/10000000000) 
    <= 6851933/8388608
  )


Reduce_Half_Pi${}_\le$

Bounds on variables:
r1 (int) $\in$ [0, 511]
x (real) $\in$ [0, 802]

assert
  -500000001 / 1000000000 <=
    (((x / (13176795/8388608)) - r1) + (542101/2500000000000000000000000))

assert
  (((x / (13176795/8388608)) - r1) - (542101/2500000000000000000000000)) <=
    500000001 / 1000000000

assert
  $\neg$(
    ((((((x - (r1 * (25735/16384))) - (r1 * (3797/67108864))) 
        - (r1 * (17453 / 17592186044416)))
       - (r1 * (12727493 / 2361183241434822606848)))
      - (x - ((r1 * $\pi$) / 2)))
     + (/ 1765573 10000000000))
    <= 18 / 100000 
  )
\end{lstlisting}

Listing \ref{code:HIEReduceVCs} shows two of the simplified exact NVCs arising from the post-conditions in Reduce\_Half\_Pi.
In both NVCs, the second and third assertions come from the third and fourth post-conditions and define how the \lstinline{x} and \lstinline{r1} variables are dependent on each other.
In both NVCs, the final assertion comes from the post-condition used to derive the NVC.
The final assertion in the first NVC asserts an upper bound on the new value of \lstinline{X} after calling Reduce\_Half\_Pi.
The final assertion in the second NVC asserts that the difference between the new value of \lstinline{X} after calling Reduce\_Half\_Pi and performing the same number of $\nicefrac{\pi}{2}$ reductions on the original value of \lstinline{X} using the exact $\pi$ is \emph{not} smaller than or equal to $\nicefrac{18}{100000}$.

\begin{lstlisting}[caption=Selected Approx\_Sin NVC, label={code:HIEApproxSinVC},basicstyle=\fontsize{7.5}{9pt}\tt,float=t]
Approx_Sin${}_\le$

Bounds on variables:
result__1 (real) $\in$ [-7639663/8388608, 3819831/4194304] -- $\sim [-0.91072, 0.91072]$
x (real) $\in$ [-6851933/8388608, 6851933/8388608]          -- $\sim [-0.81681,0.81681]$

NVC:
assert
  (abs(x) <= 1 / 67108864 $\implies$ (x = result__1))

assert
  $\neg$ (abs(x) <= 1 / 67108864) $\implies$
    (((x*((((((-3350387 / 17179869184)*(x*x)) + (4473217 / 536870912))*(x*x))
      - (349525 / 2097152))*(x*x))) + x) - (2263189 / 50000000000000))
    <= result__1
    $\wedge$
    result__1 <= 
    (((x*((((((-3350387 / 17179869184)*(x*x)) + (4473217 / 536870912))*(x*x)) 
      - (349525 / 2097152))*(x*x))) + x) + (2263189 / 50000000000000))

assert
  $\neg$( result__1 - sin(x) <= 58 / 1000000000 )
\end{lstlisting}

The exact NVC in Listing \ref{code:HIEApproxSinVC} comes from the final post-condition from the Approx\_Sin procedure in Listing \ref{code:HIEApproxSpec}.
The second and third assertions specify a dependency on \lstinline{x} and \lstinline{result__1}.
There are two assertions here due to the two if-then-else branches in the implementation of Approx\_Sin in Listing \ref{code:HIEApproxImpl}.
The final assertion specifies that the difference between the result of Approx\_Sin for \lstinline{x} and the value of the exact sine function for \lstinline{x} is not smaller than or equal to $\nicefrac{58}{1000000000}$.

\begin{lstlisting}[caption=Selected Sin NVC, label={code:HIESinVCs},basicstyle=\fontsize{7.5}{9pt}\tt,float=t]
Sin${}_\ge$

Bounds on variables:
finalresult1 (real) $\in$ [-1, 1]
o (real) $\in$ [-802, 802]
r1 (int) $\in$ [0, 511]
result__1 (real) $\in$ [-1, 1]
x (real) $\in$ [-802, 802]
y (real) $\in$ [-6851933/8388608, 6851933/8388608] 
                -- $6851933/8388608 = \texttt{Max\_Red\_Trig\_Arg} - 0.26 * \texttt{pi}$

NVC:
assert-1 x < 0.0    -> o = -x
assert-2 $\neg$(x < 0.0) -> o =  x
assert-3 -500000001 / 1000000000 <= ((o / (13176795 / 8388608)) - r1)
                                    + (542101 / 2500000000000000000000000)
assert-4 ((o / (13176795 / 8388608)) - r1) 
         - (542101 / 2500000000000000000000000) <= 500000001 / 1000000000
assert-5 -18.0 / 100000.0 <= (y + (o + (r1 * Pi / 2.0)))
assert-6 (y + (o + (r1 * Pi / 2.0))) <= 18.0 / 100000.0
assert-7 mod r1 4 <= 3.0
assert-8 
  (mod r1 4 <= 0.0) $\vee$ ($\neg$(mod r1 4 <= 0.0) $\wedge$ (mod r1 4 == 2.0)) ->
    -58.0 / 1000000000 <= result__1 - (sin y) $\wedge$
    result__1 - (sin y) <= 58.0 / 1000000000
assert-9
  $\neg$($\neg$(mod r1 4 <= 0.0) -> mod r1 4 = 2.0) ->
    -14.0 / 100000000 <= result__1 - (cos y) $\wedge$
    result__1 - (cos y) <= 14.0 / 100000000
assert-10
  x < 0 ->
    mod r1 4 <= 2 -> finalresult1 = result__1 $\wedge$
    $\neg$(mod r1 4 <= 2) -> finalresult1 = -result__1
assert-11
  $\neg$(x < 0) ->
    mod r1 4 <= 2 -> finalresult1 = -result__1 $\wedge$
    $\neg$(mod r1 4 <= 2) -> finalresult1 = result__1
assert-12 $\neg$(-19 / 100000 <= (finalresult1 - (sin x)))
\end{lstlisting}
Finally, the NVC in Listing \ref{code:HIESinVCs} comes from the first post-condition in the procedure Sin in Listing~\ref{code:HIESinImpl}.

This NVC is interesting since the implementation of the Sin procedure depends on the other procedures we have discussed, which results in the derived NVC including assertions derived from specifications of these other procedures.
Assertions 1--2 come from the if statement defining \lstinline{Y}.
Assertions 3--6 come from the Reduce\_Half\_Pi post-conditions as a consequence of calling Reduce\_Half\_Pi in Listing \ref{code:HIESinImpl}.
Assertion 7 comes from the \lstinline{Quadrant} subtype defined in Listing \ref{code:HIEReduceSpec}
Assertions 8--9 contain the final two Approx\_Sin/Approx\_Cos post-conditions as well as corresponding to one of the if-then-else branches after the call to Reduce\_Half\_Pi.
Assertions 10--11 correspond to the different paths from the final if-then-else. 
The final assertion clearly comes from the first post-condition in Listing \ref{code:HIESinImpl}.

\section{Benchmarking the Proving Process}\label{sec:benchmarks}

Tables \ref{table:benchmarks} shows
the performance of our implementation of the proving process on the verification examples described earlier.
``VC processing'' is the time it takes PropaFP to process the NVCs generated by GNATprove/Why3. 
The remaining columns in Table~\ref{table:benchmarks} show the performance on the following provers applied to the resulting simplified exact NVCs:
\begin{itemize}
  \item dReal v4.21.06.2 \cite{gao_dreal_2013} -- solver using numerical branch-and-prune methods.
  \item MetiTarski v2.4 \cite{akbarpour_metitarski_2010} -- symbolic theorem prover deciding real inequalities via cylindrical algebraic decomposition (CAD).
  \item LPPaver v0.0.1 \cite{noauthor_rasheedjalppaver_nodate} -- our prototype prover that uses methods similar to dReal.
\end{itemize}



In Table \ref{table:benchmarks}, n/s means the NVC contains some operation or number that is not supported by the prover (dReal does not support very large integers) 
while g/u means that the prover gave up.



\begin{table}[tb]
  \caption{Proving Process on Described Examples}
  \label{table:benchmarks}
  \centering
  \begin{tabular}{l@{\kern1em}r@{\kern1em}r@{\kern1em}r@{\kern1em}r}
  \toprule 
  VC                         & VC Processing & dReal     & MetiTarski  & LPPaver \\
  \midrule
  My\_Machine\_Rounding${}_\ge$    & 0.53s   & n/s       & n/s         & 0.47s \\
  My\_Machine\_Rounding${}_\le$    & 0.56s   & n/s       & n/s         & 0.42s \\
  Reduce\_Half\_Pi\_X${}_\ge$      & 1.76s   & n/s       & 0.07s       & 0.35s \\
  Reduce\_Half\_Pi\_X${}_\le$      & 1.77s   & n/s       & 0.04s       & 0.33s \\
  Reduce\_Half\_Pi${}_\ge$         & 65.02s  & n/s       & g/u         & 0.02s \\
  Reduce\_Half\_Pi${}_\le$         & 61.32s  & n/s       & g/u         & 0.01s \\
  Approx\_Sin${}_\ge$              & 1.85s   & 1m08.95s  & 0.17s       & 5.63s \\
  Approx\_Sin${}_\le$              & 1.86s   & 1m06.16s  & 0.15s       & 5.61s \\
  Approx\_Cos${}_\ge$              & 0.95s   & 3.28s     & 0.05s       & 1.83s \\
  Approx\_Cos${}_\le$              & 1.00s   & 1.53s     & 0.04s       & 1.50s \\
  Sin${}_\ge$                      & 1.29s   & n/s       & n/s         & 6m34.62s \\ 
  Sin${}_\le$                      & 1.30s   & n/s       & n/s         & 6m29.8s \\ 
  Taylor\_Sin                      & 2.04s   & 0.01s     & 0.14s       & 0.06s \\
  Taylor\_Sin\_Double              & 2.07s   & n/s       & 0.11s       & 0.05s \\
  Taylor\_Sin\_P                   & 2.05s   & 0.01s     & 0.14s       & 0.06s \\
  SinSin                           & 0.53s   & 3m19.81s  & g/u         & 8.20s \\
  Heron\_Init                      & 2.01s   & 0.00s     & 0.07s       & 0.01s \\
  Heron\_Pres                      & 3.05s   & 5m06.14s  & g/u         & 1m19.99s \\
  \bottomrule
  \end{tabular}
\end{table}


All of the NVCs were solved by at least one of the provers in a reasonable time frame.
VC processing takes, at most, a few seconds for most of the NVCs.
For the Reduce\_Half\_Pi\{${}_\ge$,${}_\le$\} NVCs, the VC processing step takes around one minute.
Most of this time was spent on the `Eliminating FP operations' step.
This is because FPTaylor takes a while to run its branch-and-bound algorithm on the file that is produced from these NVCs, particularly due to the use of $\pi$ in a non-trivial formula which can be seen in the last two post-conditions in Listing~\ref{code:HIEReduceSpec}.

Some of the NVCs could only be decided by LPPaver due to the following:
\begin{itemize}
  \item The My\_Machine\_Rounding NVC contains integer rounding with ties going away from zero.
  \begin{itemize}
    \item dReal does not support integer rounding.
    \item MetiTarski does not support the rounding mode specified in this NVC.
  \end{itemize}
  \item After our proving process, the bound on the \textbf{maximum rounding error} computed by FPTaylor in the Reduce\_Half\_Pi and the Taylor\_Sin\_Double NVCs are very small.
  \begin{itemize}
    \item This number is represented as a rational number in the exact NVC, and the denominator is outside the range of integers supported by dReal.
  \end{itemize} 
  \item The Reduce\_Half\_Pi\{${}_\ge$,${}_\le$\} NVCs have a tight bound.
  \begin{itemize}
    \item Slightly loosening the specification bounds from 0.00018 to 0.0002 allows MetiTarski to verify this.
    \begin{itemize}
      \item After this loosening, the Sin\{${}_\ge$,${}_\le$\} would need to be loosened from 0.00019 to 0.00021 due to the increased \textbf{subprogram specification error}.
    \end{itemize} 
  \end{itemize}
  \item The Sin NVCs contain integer rounding with ties going to the nearest even integer and uses the modulus operator.
  \begin{itemize}
    \item dReal does not support integer rounding.
    \item MetiTarski does not support the modulus operator.
  \end{itemize}
\end{itemize}

\subsection{Effect of Specification Bounds on Proving Times}
For numerical provers, the tightness of the specification bound is often correlated with the time it takes for a prover to decide a VC arising from said specification.
This is not normally the case for symbolic solvers, however, a VC arising from a specification that a symbolic solver could not decide may become decidable with a looser bound on the specification.
We illustrate this in Table~\ref{table:benchmarks-spec}.
The `Bound' column states the specification bound for the NVC.

\begin{table}[tb]
  \caption{Effect of Specification Bound on Proving Time}
  \label{table:benchmarks-spec}
  \centering
  \begin{tabular}{l@{\kern1em}r@{\kern1em}r@{\kern1em}r@{\kern1em}r@{\kern1em}r}
  \toprule 
  VC                                   & Bound        & VC Processing & dReal     & MetiTarski  & LPPaver \\
  \midrule
  Approx\_Sin${}_\ge$                 & -0.000000058  & 1.85s         & 1m08.95s  & 0.17s       & 5.63s \\
  Approx\_Sin${}_\ge$                 & -0.000000075  & 1.84s         & 28.27s    & 0.16s       & 3.72s \\
  Approx\_Sin${}_\ge$                 & -0.0000001    & 1.82s         & 15.07s    & 0.13s       & 2.74s \\
  Approx\_Sin${}_\ge$                 & -0.00001      & 1.87s         & 0.09s     & 0.15s       & 0.25s \\[1ex]
  Approx\_Sin${}_\le$                 & 0.000000058   & 1.86s         & 1m06.16s  & 0.15s       & 5.61s \\
  Approx\_Sin${}_\le$                 & 0.000000075   & 1.85s         & 28.73s    & 0.16s       & 3.72s \\
  Approx\_Sin${}_\le$                 & 0.0000001     & 1.86s         & 15.42s    & 0.15s       & 2.69s \\
  Approx\_Sin${}_\le$                 & 0.00001       & 1.85s         & 0.09s     & 0.15s       & 0.25s \\[1ex]
  Approx\_Cos${}_\ge$                 & -0.00000014   & 0.95s         & 3.28s     & 0.05s       & 1.83s \\
  Approx\_Cos${}_\ge$                 & -0.0000005    & 0.96s         & 0.31s     & 0.05s       & 0.61s \\
  Approx\_Cos${}_\ge$                 & -0.000001     & 0.93s         & 0.14s     & 0.05s       & 0.54s \\
  Approx\_Cos${}_\ge$                 & -0.0001       & 0.96s         & 0.00s     & 0.06s       & 0.07s \\[1ex]
  Approx\_Cos${}_\le$                 & 0.00000014    & 1.00s         & 1.53s     & 0.04s       & 1.50s \\
  Approx\_Cos${}_\le$                 & 0.0000005     & 0.94s         & 0.29s     & 0.04s       & 0.60s \\
  Approx\_Cos${}_\le$                 & 0.000001      & 0.96s         & 0.14s     & 0.04s       & 0.51s \\
  Approx\_Cos${}_\le$                 & 0.0001        & 0.94s         & 0.00s     & 0.04s       & 0.06s \\[1ex]
  SinSin                              & 0.00051778    & 0.53s         & 3m14.06s  & g/u         & 8.20s \\
  SinSin                              & 0.00052       & 0.57s         & 0.12s     & g/u         & 5.25s \\
  SinSin                              & 0.001         & 0.53s         & 0.02s     & g/u         & 1.34s \\
  SinSin                              & 0.01          & 0.53s         & 0.00s     & g/u         & 0.33s \\
  \bottomrule
  \end{tabular}
\end{table}

Table \ref{table:benchmarks-spec} shows how, in all of our examples, a looser bound results in quicker proving times for the tested numerical provers.
In some cases, this improvement can be significant, as seen with the `SinSin' NVCs.
The proving time for symbolic provers does not improve with looser bounds. 

\subsection{Counter-examples}
When writing specifications, it is not uncommon for a programmer to make a mistake in the specification by, for example, using wrong mathematical operations, setting too tight a bound for a specification, and so on.
When this occurs, it would be useful for a programmer to receive a counter-example for their specification.

Our proving process supports producing counter-examples and, with a custom Why3 driver, these counter-examples can be reported back to Why3, which will send the counter-examples to the programmer's IDE.
It should be understood that counter-examples produced by the proving process are \emph{potential} counter-examples, since `simplified exact' NVCs are weakened versions of original NVCs.
Nevertheless, these \emph{potential} counter-examples can still be \emph{actual} counter-examples and would be useful for a programmer to have.

To demonstrate how the proving process can produce counter-examples, we modify our \lstinline{Taylor_Sin} example, introducing three different mistakes which a programmer may feasibly make:
\begin{enumerate}
  \item Replace the \lstinline{-} with \lstinline{+} in the \lstinline{Taylor_Sin} implementation in Listing~\ref{code:adaSineImpl}.
  \item Invert the inequality in the \lstinline{Taylor_Sin} post-condition in Listing~\ref{code:sineSpec}.
  \item Make our specification bound slightly tighter than the \textbf{maximum model error} + \textbf{maximum rounding error} + \textbf{rounding analysis cushion} in the post-condition from Listing \ref{code:sineSpec}, changing the value of the right hand side of the inequality in the post-condition from 0.00025889 to 0.00025887.
\end{enumerate}
These three `mistakes' are referred to as Taylor\_Sin\_Plus, Taylor\_Sin\_Swap, and Taylor\_Sin\_Tight, respectively, in Table~\ref{table:counter-examples}.

If a specification is incorrect, the resulting NVC must be true or `sat'.
dReal would report a `delta-sat' result, which means the given file was sat with a configurable tolerance, which we set to $1^{-100}$.
This makes models produced by dReal a \emph{potential} model for the NVC.
Models produced by LPPaver are actual models for the given NVC, but for files produced by the proving process, these should still be thought of as \emph{potential} counter-examples due to the weakening of the NVC.
The computed potential counter-examples shown in Table~\ref{table:counter-examples} are all actual counter-examples except those for Taylor\_Sin\_Tight.

\begin{table}[t]
  \caption{Proving Process on Described Counter-examples}
  \label{table:counter-examples}
  \centering
  \begin{tabular}{l@{\kern-2em}r@{\kern1em}r@{\kern1em}l@{\kern1em}r@{\kern1em}l}
  \toprule
  VC                 & VC Processing & dReal & CE & LPPaver & CE \\
  \midrule
  Taylor\_Sin\_Plus  & 2.05s         & 0.00s &  $x=-0.166\ldots$  & 0.02s & $x=-0.5$ \\
  Taylor\_Sin\_Swap  & 2.05s         & 0.00s &  $x=0.166\ldots$   & 0.03s & $x=0.499\ldots$\\
  Taylor\_Sin\_Tight & 2.1s          & 0.00s &  $x=0.499\ldots $  & 0.03s & $x=0.499\ldots$\\
  \bottomrule
  \end{tabular}
\end{table}


\section{Conclusion}\label{sec:conclusion}

\paragraph{Summary}
In this paper, we have presented an automated proving process for deciding VCs that arise in the verification of floating-point programs with a strong functional specification.
The proving process combines several existing techniques and tools in its steps (cf., Fig.~\ref{fig:gnatprove}):

\begin{enumerate}
  \item Why3 reads a program+specification and produces NVCs (Negated VCs). 
  \item PropaFP processes the NVCs as follows:
  \begin{enumerate}
    \item Simplify the NVC using simple symbolic rules and interval evaluation.
    \item Derive bounds for all variables in the NVC, interleaving with (a).
    \item Derive bounds for rounding errors in expressions with FP operations.
    \item Using these bounds, safely replace FP operations with exact operations.
    \item Repeat the simplification steps (a--b).
  \end{enumerate}
  \item Apply nonlinear real provers on the processed NVCs to either prove them or get \emph{potential} counter-examples.
\end{enumerate}

This proving process should, in principle, work with other tools and languages than Why3 and SPARK, as long as one can generate NVCs similar to those generated by GNATprove. 

We demonstrated our proving process on three examples of increasing complexity, featuring loops, real-integer interactions and subprogram calls.
Notably, we have contributed the first fully verified SPARK implementations of the sine and square root functions. 
The examples demonstrate an improvement on the state-of-the-art in the power of automated FP software verification.


Table \ref{table:benchmarks} indicates that our proving process can automatically and fairly quickly decide certain VCs that are currently considered difficult.
Table \ref{table:benchmarks-spec} demonstrates how the process speeds up when using looser bounds in specifications.
Table \ref{table:counter-examples} shows that our proving process can quickly find potential, often even actual, counter-examples for a range of common incorrect specifications.

Our examples may be used as a suite for benchmarking future FP verification approaches, and our resulting NVCs can be used to benchmarks future nonlinear real provers.

\subsection*{Future work}

We conclude with thoughts on how our process could be further improved.

\paragraph*{Executable exact real specifications.}
We plan to make specifications containing functions such as $\sqrt{\cdot}$ executable via high-accuracy interval arithmetic, allowing the developer or IDE to check whether the {suggested} counter-examples are real.

\paragraph*{Improving the provers.}
We would like the provers we used in this paper to be improved in some ways.
Ideally, the provers will be able to decide all of our examples.
Support for integer rounding could be added to dReal, using methods similar to those used in LPPaver.
It should also not be difficult to add support for larger integers in dReal.
Support for  both integer rounding and the modulus operator could be added to MetiTarski.
Adding these features will allow both dReal and MetiTarski to have an attempt at deciding all of our examples.

\paragraph*{Why3 integration.}
Our VC processing could be integrated into Why3.
This would include the symbolic processing, bound derivation, and floating-point elimination.
As Why3 transformations, the VC processing steps would be more accessible for users who are familiar with Why3.
Also, the proving process would thus become easily available to the many tools that support Why3.

\paragraph*{Support function calls.}
Having to manually translate functions into procedures is undesirable.
Support for function calls could be added, e.g.,
by a Why3 transformation that translates functions into procedures.

\paragraph*{Use Abstract Interpretation.}
We currently derive bounds for variables using our own iterative process similar to Abstract Interpretation over the interval domain.
It would be interesting to see if the proving process would improve if we use an established Abstract Interpretation implementation to derive bounds.
If nothing else, such change would reduce the amount of new code that a user would need to trust.


\paragraph*{Verified implementation.}

We would like to formally verify some elements of our process to ensure that the transformation steps are performed correctly.
As PropaFP and LPPaver utilise Haskell and AERN2 \cite{noauthor_aern2_nodate}, rewriting these tools in Coq with coq-aern \cite{konecny_axiomatic_2021} may be a feasible verification route.


%
%
\bibliographystyle{splncs04}
\bibliography{mybibliography}

\end{document}